\documentclass[iop]{emulateapj}
\newcommand{\kms}{{\rm{km\,s^{-1}}}}
\newcommand{\hi}{\ion{H}{1}}

\slugcomment{ApJ Accepted 6 February 2018}

\shorttitle{M82 Minor Axis Kinematics}
\shortauthors{Martini et al.}

\begin{document}

\title{\hi\ Kinematics Along The Minor Axis of M82}  

\author{Paul Martini\altaffilmark{1,2}, 
Adam K. Leroy\altaffilmark{1}, 
Jeffrey G. Mangum\altaffilmark{3}, 
Alberto Bolatto\altaffilmark{4}, 
Katie M. Keating\altaffilmark{5},
Karin Sandstrom\altaffilmark{6},
Fabian Walter\altaffilmark{7}}

\altaffiltext{1}{Department of Astronomy, The Ohio State University, Columbus, OH 43210, USA, martini.10@osu.edu, leroy.42@osu.edu} 
\altaffiltext{2}{Center for Cosmology and Astroparticle Physics, The Ohio State University, Columbus, OH 43210, USA} 
\altaffiltext{3}{National Radio Astronomy Observatory, 520 Edgemont Road, Charlottesville, VA 22903, USA}
\altaffiltext{4}{Department of Astronomy, University of Maryland, College Park, MD 20742, USA} 
\altaffiltext{5}{Rincon Research Corporation, 101 North Wilmot Road, Suite 101, Tucson, AZ 85711, USA} 
\altaffiltext{6}{Department of Physics, University of California, San Diego, CA 92093, USA}
\altaffiltext{7}{Max-Planck-Institute f\"ur Astronomie, K\"onigstuhl 17, D-69117 Heidelberg, Germany} 

\begin{abstract}
M82 is one of the best studied starburst galaxies in the local universe, and is consequently a benchmark for studying star formation feedback at both low and high redshift. We present new VLA \hi\ observations that reveal the cold gas kinematics along the minor axis in unprecedented detail. This includes the detection of \hi\ up to 10 kpc along the minor axis toward the South and beyond 5 kpc to the North. A surprising aspect of these observations is that the line-of-sight \hi\ velocity decreases substantially from about $120\, \kms$ to $50\, \kms$ from 1.5 to 10 kpc off the midplane. The velocity profile is not consistent with the \hi\ gas cooling from the hot wind. We demonstrate that the velocity decrease is substantially greater than the deceleration expected from gravitational forces alone. If the \hi\ consists of a continuous population of cold clouds, some additional drag force must be present, and the magnitude of the drag force places a joint constraint on the ratio of the ambient medium to the typical cloud size and density. We also show that the \hi\ kinematics are inconsistent with a simple conical outflow centered on the nucleus, but instead require the more widespread launch of the \hi\ over the $\sim 1$ kpc extent of the starburst region. Regardless of the launch mechanism for the \hi\ gas, the observed velocity decrease along the minor axis is sufficiently great that the \hi\ may not escape the halo of M82. We estimate the \hi\ outflow rate is much less than 1 M$_{\odot}$ yr$^{-1}$ at 10 kpc off the midplane.  
\end{abstract}


\section{Introduction}

Galaxies with substantial star formation rates often have large-scale winds that represent material flowing out of the galaxy along the minor axis \citep[e.g.\ ][]{heckman90,veilleux05}. These winds are most likely driven by young, massive stars, which have substantial stellar winds and explode as core-collapse supernovae. The material in these winds can entrain all phases of the interstellar medium, and is expected to be preferentially metal-enriched by the nucleosynthetic products in supernova ejecta. If the material in these winds escapes the galaxy, the winds can enrich the circumgalactic and intergalactic medium \citep[e.g.\ ][]{borthakur13,werk16} and  regulate the growth and metal-enrichment histories of galactic disks \citep[e.g.\ ][]{oppenheimer08,peeples11}. Observational evidence of the removal of metal-encirched material includes the metallicity of the intergalactic medium, the mass-metallicity relationship, and the sizes and luminosities of galactic disks.

M82 is arguably one of the best-studied galaxies with a starburst-driven wind \citep{lynds63}. Its two key properties are its proximity at only $D = 3.63$ Mpc \citep{freedman94,gerke11} in the M81 group and that the galaxy is observed nearly edge on, which makes it particularly well-suited to observe the outflow along the minor axis. Studies of this galaxy across the electromagnetic spectrum have detected multiple phases of the wind moving at multiple velocities, ranging from plasma temperatures of $30 - 80$ million Kelvin at X-ray energies \citep{schaaf89,strickland09} to warm, ionized gas traced by visible-wavelength features such as H$\alpha$ \citep{mckeith95,westmoquette09} to atomic \hi\ and molecular gas at temperatures of 100 K or less \citep{seaquist01,walter02,salak13,leroy15}. 
 
Observations of visible emission lines along the minor axis showed line splitting in position velocity diagrams \citep{axon78,amirkhanyan82}, and subsequent Fabry-Perot data exhibited good evidence for a large-scale, biconical outflow \citep{bland88}. The region with double-peaked emission lines begins about $300$ pc from the midplane and has been detected to about 1 kpc \citep{heckman90}. \citet{mckeith95} derived an inclination of $80^\circ$ for the galactic disk, and a cone opening angle of $30^\circ$ (semi-angle $15^\circ$) at $> 1$ kpc. Their model starts as a cylindrical flow and transitions to a bicone at about $300$ pc, and  then radially accelerates to a terminal velocity of $600\ \kms$ on the surface of the bicone. The material inside the bicone is the much hotter fluid observed at X-ray energies and predicted to have a terminal velocity several times greater than the $\sim 500\,\kms$ escape velocity \citep{strickland09}. 

The line of sight velocities of $120 - 140\ \kms$ relative to systemic for the cold atomic and molecular gas traced by \hi\ and CO are substantially slower than the emission traced at shorter wavelengths. The cold gas could be entrained in the hot gas, although simulations suggest that only a small fraction of the cold gas should survive a significant amount of acceleration \citep{scannapieco15,schneider17}, except perhaps if the clouds are supported by magnetic fields \citep{mccourt15}. Another possibility is that the winds are driven by radiation pressure through momentum transfer from starlight to dust grains embedded in the gas \citep{murray05,murray10,andrews11}. Recent hydrodynamic simulations by \citet{zhang17} suggest that clouds accelerated by the radiation field last significantly longer than clouds entrained in the hot outflow. Both the atomic and molecular gas show double-peaked lines that suggest the hot, conical outflow is bounded by a sheath of cooler material \citep{walter02}. \citet{leroy15} conclude that the cold \hi\ and CO help to confine the hot outflow. The best evidence for this connection begins approximately 1.5 kpc from the disk. At these distances the \hi\ and CO exhibit double-peaked profiles indicative of a bicone, a velocity gradient consistent with the outflow, and spatial coincidence with features at other wavelengths \citep[see Figure 16 in][]{leroy15}. 

In this paper we present new, wide-field \hi\ observations of M82 obtained with the Karl G. Jansky Very Large Array (VLA) combined with previous observations with the 100-m Robert C. Byrd Green Bank Telescope (GBT). These observations show the \hi\ intensity and kinematics over approximately $20\arcmin \times 20\arcmin$ with a resolution of $24\arcsec$. Based on the distance $D = 3.63$ Mpc, $1 {\rm kpc} \approx 1'$. These observations therefore trace the \hi\ outflow up to a projected distance of $10$ kpc along the minor axis. The next section describes our observations in further detail, and the following section presents a mass model for M82. We use this mass model in \S\ref{sec:an} to explore various explanations for the minor axis kinematics, and summarize our results in \S\ref{sec:sum}. Except where otherwise noted, we adopt an inclination of $80^\circ$ for the disk of M82 \citep{lynds63,mckeith93}. 

\section{Observations} \label{sec:obs}

\begin{figure*}[ht]
\plotone{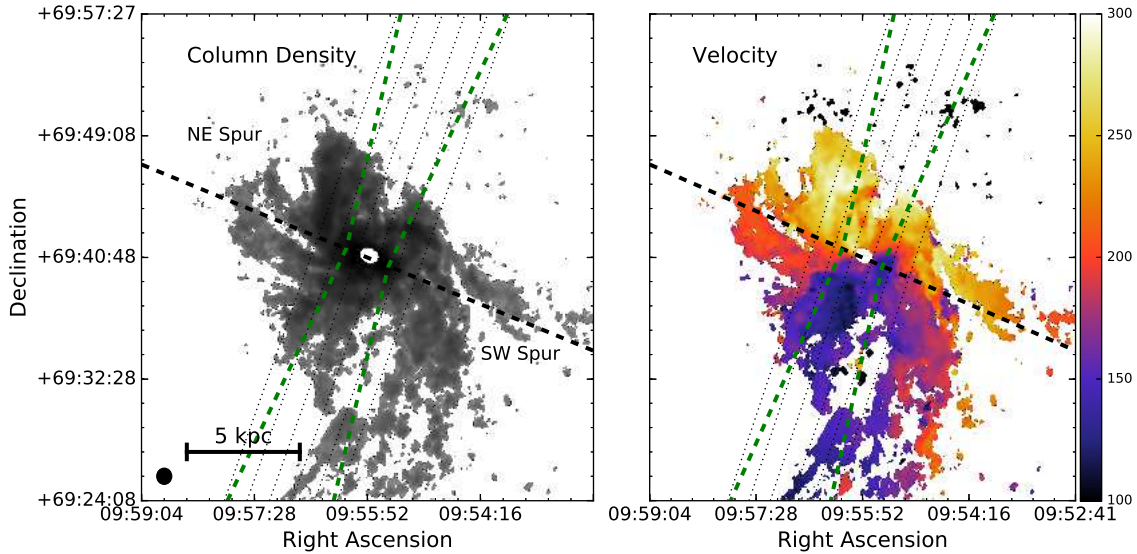}
\caption{Log of the M82 \hi\ column density ({\it left}) and velocity field ({\it right}). The velocity scale in $\kms$ is shown on the right colorbar. The major axis is oriented at $PA = 67^\circ$ ({\it black, dashed line}). The outline of our biconical frustum model ({\it green, dashed line}) is oriented along the minor axis with an opening angle of $\theta = 20^\circ$ (see \S~\ref{sec:frustum}). Also shown are the outlines of various 1 kpc wide slices along the minor axis used to compute position-velocity diagrams and a scale bar with a projected length of 5 kpc. The absence of \hi\ at the center position is because there is \hi\ absorption of the bright continuum source. \label{fig:intvel}}
\end{figure*}

\begin{figure*}[ht]
\plotone{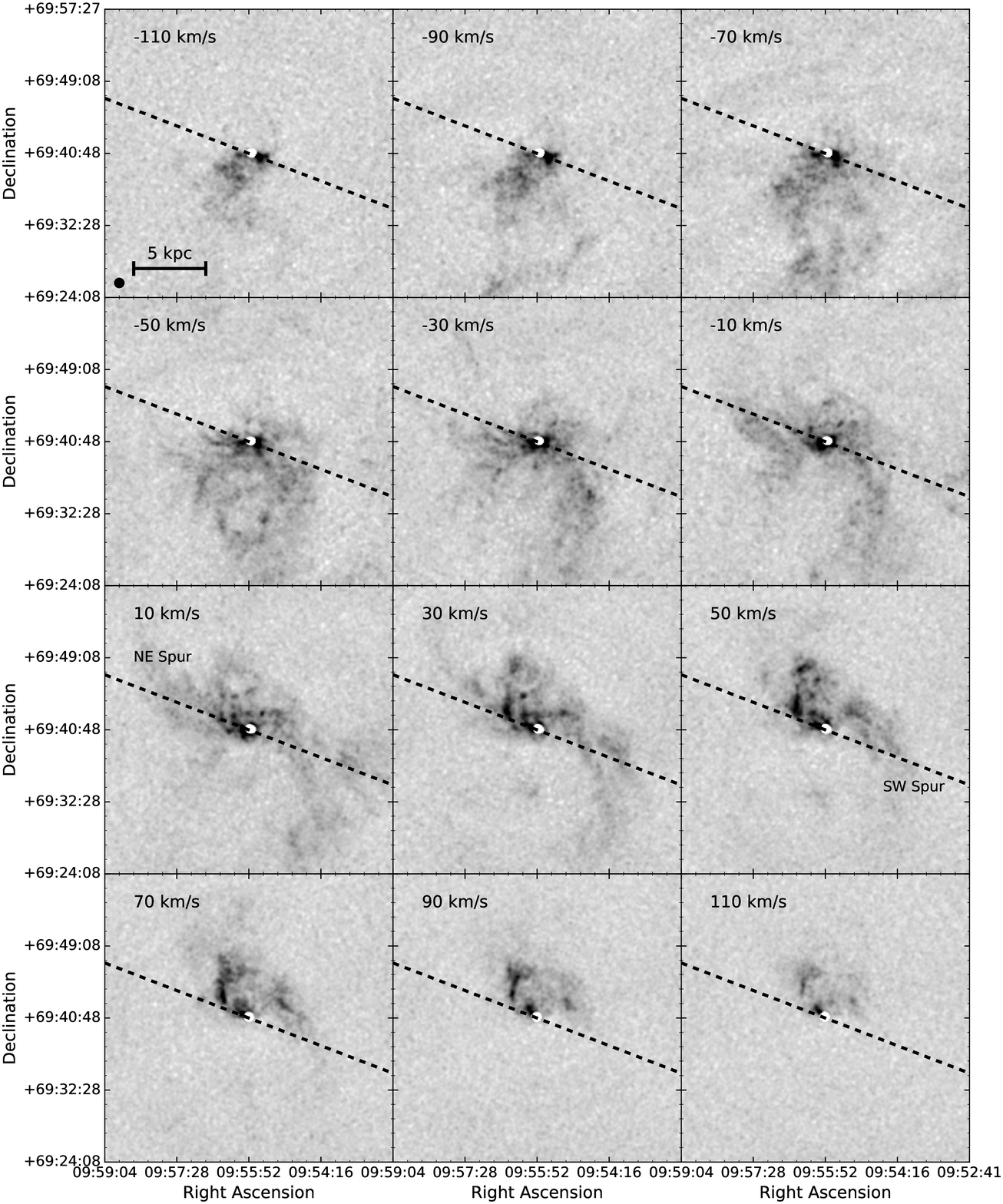} 
\caption{M82 \hi\ intensity in twelve velocity channels after continuum subtraction. The central white area is due to \hi\ absorption toward the center of the galaxy. The velocity for each channel is relative to the systemic velocity of M82. The major axis is oriented at $PA = 67^\circ$ ({\it black, dashed line}). The projected size of 5 kpc and the $24''$ FWHM beam diameter are shown in the upper left panel. \label{fig:channels}}
\end{figure*}

We observed M82 in the L band with the VLA in the B, C, and D configurations in 2015 and 2016, and combined these data with L band observations from the GBT obtained between 2003 and 2009. The VLA observations were centered on M82 at $\alpha = 9$:55:52.72, $\delta = +69$:40:45.7 (J2000), and on-source integration times in the B, C, and D configurations were $\sim 378$, $294$, and $168$ minutes, respectively. The GBT observations consist of 40 sessions that include the M81/M82 and NGC2403 galaxy groups. The GBT observations were used to construct an $8.7^\circ \times 21.3^\circ$ map centered at $\alpha = 8$:40:37.0, $\delta = +69$:17:16 (J2000) and have a total integration time of approximately 187 hours. This GBT map includes data previously described in \citet{chynoweth08,chynoweth09,chynoweth11}. 

\subsection{VLA Data Processing} 

We set the VLA correlator to observe the full, polarized L-band continuum with 1 MHz channels across the full range $\sim 1{-}2$~GHz. We also configured spectral windows to target the \hi\ and OH lines at higher spectral resolution. For the \hi\ line, which is the focus of this paper, we used 2048 3.906 kHz channels centered on the \hi\ line at the velocity of M82. For the OH lines, we used a slightly coarser channel width of 7.812 kHz. We calibrated the amplitude, bandpass, and phase with measurements of $1331+305$ (3C\,286; amplitude and bandpass; flux density = 15.0\,Jy at 1.4\,GHz) and J$0921+6215$ (B-configuration phase; flux density = $1.186\pm0.003$\,Jy/beam at 1.55\,GHz), and J$0841+7053$ (C- and D-configuration phase; flux density = $3.340\pm0.018$\,Jy/beam at 1.57\,GHz), respectively, using standard techniques. This observation strategy mirrored that used successfully for THINGS, VLA ANGST, and LITTLE THINGS \citep{walter08,hunter12,ott12}. While M82 was extensively observed by the VLA in the 1980s \citep{yun93,yun94}, the current observations are more sensitive, have higher velocity resolution, and have much better dynamic range due to upgrades from the early 1990s to late 2000s that improved sensitivity and bandwidth. The later is particularly important, given M82's substantial continuum emission. 

As the continuum emission from M82 is quite strong, we self-calibrated all measurements. The reduction in phase residual following self-calibration resulted in a factor of three to five improvement in peak signal-to-noise for all measurements. After self-calibration, we split out the spectral window containing \hi, binning the data to have $8$s integrations. We identified line free channels from the integrated spectrum of the $u-v$ data. We then fit and subtracted a first order polynomial from each visibility measurement. We combined all of the continuum-subtracted visibility data for all four observations, in the process binning the data to a final channel width of $5$~km~s$^{-1}$.

We imaged the data in stages to account for the substantial absorption in the center of M82, . First, we created a high resolution image with Briggs weighting parameter ${\tt robust} =0$ and no $u-v$ taper. This image had a beam size of $\sim5\arcsec$. The surface brightness sensitivity of this image was too poor to study the outflow in detail, but yielded a good initial model of the inner part of the galaxy. We cleaned in interactive mode until the maximum residual appeared comparable to the noise in the image. We then imaged the data again, beginning from the model output by the previous run, and this time used ${\tt robust} =0.5$ weighting to emphasize surface brightness sensitivity slightly more. Again, we cleaned until the maximum residual was comparable to the noise. Then, starting from the model output by that imaging run, we imaged the data with an $18\arcsec$ $u-v$ taper. This dramatically improved the surface brightness sensitivity of the data. Because this step began with the previous, higher-resolution model, we achieved a substantial improvement in surface brightness sensitivity without significant, negative impact due to the strong absorption in the center. We cleaned this image until the maximum residual again resembled the noise of the image and smoothed the output image to have a round, $24\arcsec$ beam (FWHM). 

\subsection{GBT Data Processing} 

The M81/M82 and NGC 2403 area was observed by moving the telescope in declination and sampling every $3\arcmin$, with an integration time of 1-3 seconds per sample (the integration time varied with the observation session). Strips of constant declination were spaced by $3\arcmin$, and the telescope was moved in right ascension to form a ‘basket weave’ pattern over the region. The GBT spectrometer was used with a bandwidth of either 12.5 or 50 MHz, depending on the observation session. The combined bandwidth for the final map is 10.5 MHz, and has a velocity range from $-890$ to $1320\,\kms$. The typical system temperature for each channel of the dual-polarization receiver was $\approx 20$ K. 

The GBT data were reduced in the standard manner using the GBTIDL and AIPS data reduction packages. Spectra were smoothed to a channel spacing of 24.4 kHz, corresponding to a velocity resolution of $5.2\,\kms$. A reference spectrum for each of the nine observation sessions was made from an observation of an emission-free region, usually from the edges of the maps. The reference spectrum was used to perform a (signal-reference)/reference calibration of each pixel. These calibrated spectra were scaled by the system temperature and corrected for atmospheric opacity and GBT efficiency. We adopted the GBT efficiency from equation one of \citet{langston07} for a zenith atmospheric opacity of $\tau_0 = 0.009$. The frequency range observed was relatively free of radio frequency interference (RFI) and less than 0.3\% of all spectra were adversely affected. The spectra exhibiting RFI were identified by tabulating the root-mean-square (RMS) noise level in channels free of neutral hydrogen emission. Spectra that showed unusually high noise across many channels were flagged and removed. The observations were then gridded using the AIPS task {\tt SDIMG}, which also averages polarizations. After amplitude calibration and gridding, a 1st-order polynomial was fit to line-free regions of the spectra and subtracted from the gridded spectra using the AIPS task {\tt IMLIN}. The effective angular resolution, determined from maps of 3C286, is $9.15 \pm 0.05\arcmin$. To convert to units of flux density, we observed the calibration source 3C286. The calibration from Kelvin to Janskys was derived by mapping 3C286 in the same way that the HI maps were produced, and the scale factor from K/Beam to Jy/Beam is $0.43 \pm 0.03$. Due to the patchwork nature of the observations, the RMS noise varies considerably across the datacube, ranging between $6-30$ mJy/beam. The average RMS noise in the final data cube is 20 mJy per 24.4 kHz channel. 

\subsection{Construction of Combined Data Products} 

The VLA-only cube still exhibits ``bowling'' and other large scale artifacts, in part due to the absence of short and zero spacing data. To deal with this, we combined the VLA cube with the GBT cube. First, we extracted a subcube centered on M82 from the larger M81 GBT survey. We then masked channels dominated by Galactic emission, carried out one additional round of first-order baseline fitting, applied the primary beam taper of the VLA data to the GBT data, and combined the VLA and GBT data with the CASA task {\tt feather}. This combination removed or suppressed most of the large scale artifacts that were present in the VLA-only cube. After the combination, we corrected the cube for the primary beam response of the VLA. The final cube has a channel width of $5\,\kms$, beam size of $24\arcsec$ (FWHM) , and covers the primary beam of the VLA out to its half power point ($\approx 32\arcmin$). Before the primary beam correction, the cube had an rms noise of $\approx 0.4$~mJy~beam$^{-1}$ at this resolution and channel width, equivalent to $\approx 4$~K noise for $\approx 1050$~K/Jy.

We converted the cube to have units of Kelvin, and also made a version of our \hi\ cube with individual channels in units of column density. To do this we used the $5$~km~s$^{-1}$ width of each channel and assumed optically thin \hi\ emission so that $N \approx 1.823 \times 10^{18}$\,cm$^{-2}$~$I_{HI}$ with $I_{HI}$ in K~$\kms$. To highlight only bright, positive emission, we also created a mask that included all regions of the cube where the intensity exceeded a signal-to-noise ratio of five over two successive channels. This corresponds to a column density limit of $\approx 7 \times 10^{19}$~cm$^{2}$ across a $10\, \kms$ velocity range. We used this mask to create integrated intensity and intensity-weighted mean velocity maps. Note that most of our analysis focuses on direct analysis of the cubes.

The integrated \hi\ intensity map and velocity field are shown in Figure~\ref{fig:intvel}, and channel maps of \hi\ intensity in select velocity channels are shown in Figure~\ref{fig:channels}. The \hi\ intensity map shows substantial \hi\ extending along the major axis for over $\pm 5$ kpc. The \hi\ emission to the left (East) shows a significant spur that extends up (North) from the disk starting at about $4-5$ kpc. This is the tidal stream seen in previous \hi\ observations  \citep{cottrell77,vanderhulst79,appleton81,yun93,yun94}. This feature is labeled "NE Spur" in Figures~\ref{fig:intvel} and \ref{fig:channels}. Our data also show some of the extensive pair of streams that extend west of the major axis, and then toward the South \citep{yun93}. This feature is labeled "SW Spur" in Figures~\ref{fig:intvel} and \ref{fig:channels}. The \hi\ emission along the minor axis extends approximately 5 kpc toward the North and 10 kpc toward the South. M82 is to the North of M81, so the southern part of M82 is closer to M81. 

\begin{figure}[ht]
\plotone{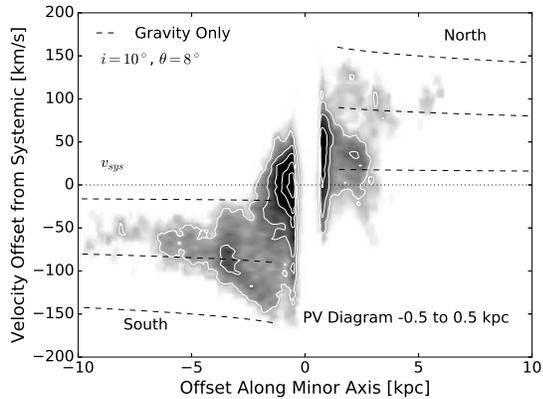} 
\caption{M82 minor axis position-velocity diagram. The grayscale and contours show the integrated \hi\ intensity within $\pm 0.5$ kpc along the minor axis as a function of projected distance from the midplane. The velocity field has been shifted to the systemic velocity of $211\,\kms$ ({\it dotted line}). \hi\ is clearly detected up to 10 kpc to the South ({\it left}) and to about 6 kpc to the North ({\it right}). Also shown are the ballistic velocities for test particles ({\it dashed lines}) for the mass model presented in \S\ref{sec:mass}. Three trajectories are shown in each direction. The central trajectory is aligned with the symmetry axis of a bicone tilted  $\theta = 10^\circ$ relative to the plane of the sky, while the other two trajectories are inclined by $\pm 8^\circ$ relative to the symmetry axis. This simple, ballistic model is not a good match to the data. \label{fig:pv1}}
\end{figure}

The \hi\ velocity field shows some evidence of rotation, similar to the molecular gas \citep{young84,yun93,leroy15}, although not as regular as the $K-$band stellar rotation curve measured by \citet{greco12}. The \hi\ rotation curve exhibits several drops in velocity \citep{sofue92}, which are most likely due to the influence of the bar \citep{telesco91} and the tidal interaction \citep{yun93}. The velocity field along the minor axis has a steep gradient within about 1 kpc, where it decreases rapidly by over $100\,\kms$ toward the South and increases rapidly by over $100\,\kms$ toward the North. This spatial extent is where the start of the outflow is observed at other wavelengths \citep{mckeith95,westmoquette09}. The amplitude of the velocity field then decreases toward the systemic velocity of M82. This is particularly prominent toward the South, where continuous, albeit filamentary, \hi\ is detected out to 10 kpc from the midplane. At this point the velocity has decreased to approximately half the value at 1 kpc. Figure~\ref{fig:pv1} shows a position-velocity diagram along the minor axis within a synthesized slit of width $\pm 0.5$ kpc relative to the center of M82. The boundaries of this slit are shown on Figure~\ref{fig:intvel}. 

\section{Mass Model} \label{sec:mass}

\begin{figure}[ht]
\plotone{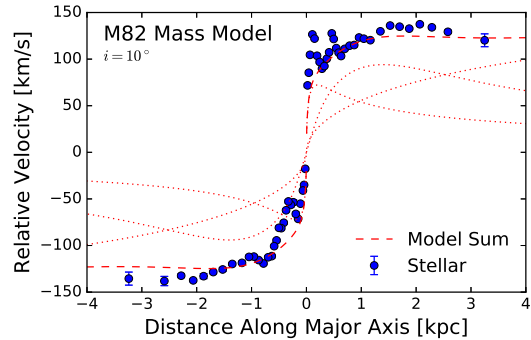} 
\caption{M82 rotation curve model and observations. The data points (with error bars) are the stellar kinematics measured by \citet{greco12} with the $2.29\mu$m CO bandhead in the near-infrared $K-$band. The lines show the rotation curve ({\it dashed}) and three individual components ({\it dotted}) calculated from the mass model described in \S\ref{sec:mass}. The bulge component is the most significant component near the center of the galaxy, the disk dominates out to several kpc, and the halo at larger radii. \label{fig:rc}}
\end{figure}

We developed a mass model for M82 based on measurements of the stellar surface photometry, stellar kinematics, and simulations of its interaction history within the M81 group. This model includes a bulge, a disk, and a dark matter halo and was constructed with the \texttt{galpy} software package \citep{bovy15}. We took the physical length scale of the bulge and disk components from the wide-field, near-infrared surface photometry of \citet{ichikawa95}. They performed a bulge-disk decomponsition and found the bulge light along the major axis is well fit by a de Vaucouleurs $R^{1/4}$ law with an effective radius of $21''$ ($K'-$band) to $29''$ ($H-$band). We adopt a physical radius of $r_e = 0.36$ kpc for this component, which we model with a Hernquist potential \citep{hernquist90} with $a = 0.2$ kpc (with the relation $a = r_e/1.8153$). \citet{ichikawa95} measured an exponential disk scale length of $40''$, and we adopt a physical disk of $r_s = 0.6$ kpc. We model the exponential disk with the \texttt{galpy} implementation of the \citet{smith15} prescription to describe an exponential disk as three Miyamoto-Nagai disks \citep{miyamoto75}. We set the vertical scale height to $h_z = 0.1$ kpc. This quantity was not calculated by \citet{ichikawa95}, but the exact choice does not appreciably impact the kinematics above the midplane. Lastly, we model the dark matter halo as a \citet[][NFW]{navarro96} profile. \hi\ observations of the M81 group show long tidal features that are good evidence of past interactions between M81 and M82, as well as between M81 and NGC 3077 \citep{cottrell77,yun93}. \citet{oehm17} performed a thorough simulation study of various scenarios for the interaction history of the M81 group and placed constraints on the dark matter profiles of the largest members. We adopt their NFW parameters of $R_{200} = 164$ kpc, $\rho_0 = 8.81 \times 10^{-3}$ M$_{\odot}$ pc$^{-3}$, and $c = 11.17$.

There are two free parameters in the mass model: the masses of the bulge and disk components. We adjust these parameters until we obtain a reasonably good match to the major axis rotation curve measured by \citet{greco12}. That rotation curve was measured from the CO $2.29\mu$m bandhead that is prominent in the near-infrared spectra of giant and supergiant stars. This rotation curve is shown in Figure~\ref{fig:rc}, along with the rotation curves of the three mass components. The mass we adopt for the bulge component is $2\times10^9$ M$_\odot$, although we note that much of this mass may not be in the form of the old stellar population characteristic of classical bulges. One reason is that there is good evidence from stellar spectroscopy that the central light distribution is dominated by young supergiants \citep{rieke93,forster01,forster03,greco12}, rather than the old stellar population typical of classical bulges. A substantial fraction of the mass in the central region is also in the form of  atomic and molecular gas  \citep{young84,walter02,leroy15}. Regardless of the exact form of the mass, this component dominates the rotation curve in the central $0.5$ kpc. We normalize the disk component to $8\times10^9$ M$_\odot$ kpc$^{-3}$, and this is the most prominent component out to about 2 kpc. The total mass of our model is in good agreement with the total dynamical mass of $\sim 10^{10} M_{\odot}$ derived by \citet{greco12}, and the range of $10^{9-10} M_{\odot}$ derived by previous studies, although those studies analyzed a smaller range of radii \citep{young84,goetz90,sofue92}. 

\section{Analysis} \label{sec:an} 

\begin{figure}[ht]
\plotone{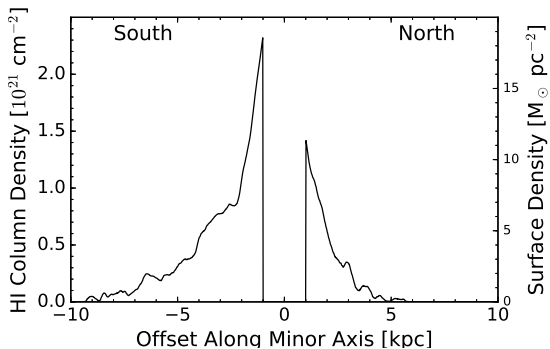} 
\caption{HI column density vs.\ distance from the midplane in the same $\pm0.5$ kpc region used to construct Figures~\ref{fig:pv1} and \ref{fig:pvgd}. The central $\pm 1$ kpc is masked as there is significant \hi\ absorption near the midplane. The right axis shows the equivalent surface density of neutral Hydrogen calculated from $8 \times 10^{-21} N_H$. \label{fig:nhi}}
\end{figure}

We detect substantial \hi\ emission along the minor axis out to at least 5 kpc to the North and 10 kpc to the South (Figure~\ref{fig:intvel}). The total \hi\ intensity is also greater toward the South than toward the North, and there is more molecular gas and dust to the South \citep{walter02,leroy15}. Figure~\ref{fig:nhi} shows the \hi\ column density as a function of distance from the midplane along the minor axis in the same $\pm 0.5$ kpc region used to construct Figure~\ref{fig:pv1}.

The spatial distribution and kinematics of the \hi\ gas are a poor match to the outflow traced by the hottest gas. \citet{strickland00} discuss how the hot wind fluid traced by their X-ray observations may accelerate cold clouds by ram pressure. \citet{strickland09} estimate the terminal velocity of the wind is $v_{\infty} = 1410 - 2240\,\kms$, which is greater than the escape velocity of $v_{esc} = 460\,\kms$ for M82. The minor axis kinematics of the warm, ionized gas traced by H$\alpha$ and other visible-wavelength lines suggest velocities of $500 - 600\,\kms$ for an $80^\circ$ disk inclination \citep{mckeith95,shopbell98}. The velocity of the warm ionized gas is consequently substantially lower than the hot wind fluid, and closer to M82's escape velocity \citep{heckman00}. The \hi\ kinematics are similar to the warm, ionized gas, as well as the colder, molecular gas traced by CO \citep{walter02,leroy15}. The similar kinematics supports the hypothesis that the warm ionized gas and the cooler atomic and molecular gas are cospatial, but they are not cospatial with the hot wind fluid detected at X-ray energies.

Further evidence that the warm ionized gas and cooler atomic and molecular gas are cospatial comes from measurements of line splitting along the minor axis that suggests both gas phases trace the edge of a bicone that surrounds the higher temperature X-ray fluid. This evidence includes the Fabry-Perot observations of \citet{bland88} and line splitting in longslit spectroscopy. \citet{heckman90} used longslit spectroscopy to derive an opening angle of $60^\circ$ (and angles of $30-40^\circ$ between the symmetry axis of the cone and the plane of the sky). \citet{mckeith95} derived a smaller cone opening angle of $30^\circ$ at $> 1$ kpc. Line splitting is also observed in the colder gas traced by \hi\ and CO. \citet{leroy15} derived a smaller cone opening angle of $13-20^\circ$ for this cold gas. The velocity spread of the minor axis \hi\ observed in our new observations is more consistent with the smaller opening angle derived by \citet{leroy15}. A possible explanation for the discrepancy between the measurements from warm, ionized gas and colder atomic gas are that the later measurements were made further from the midplane, where the outflow may be more columnated.

\begin{figure*}[ht]
\plotone{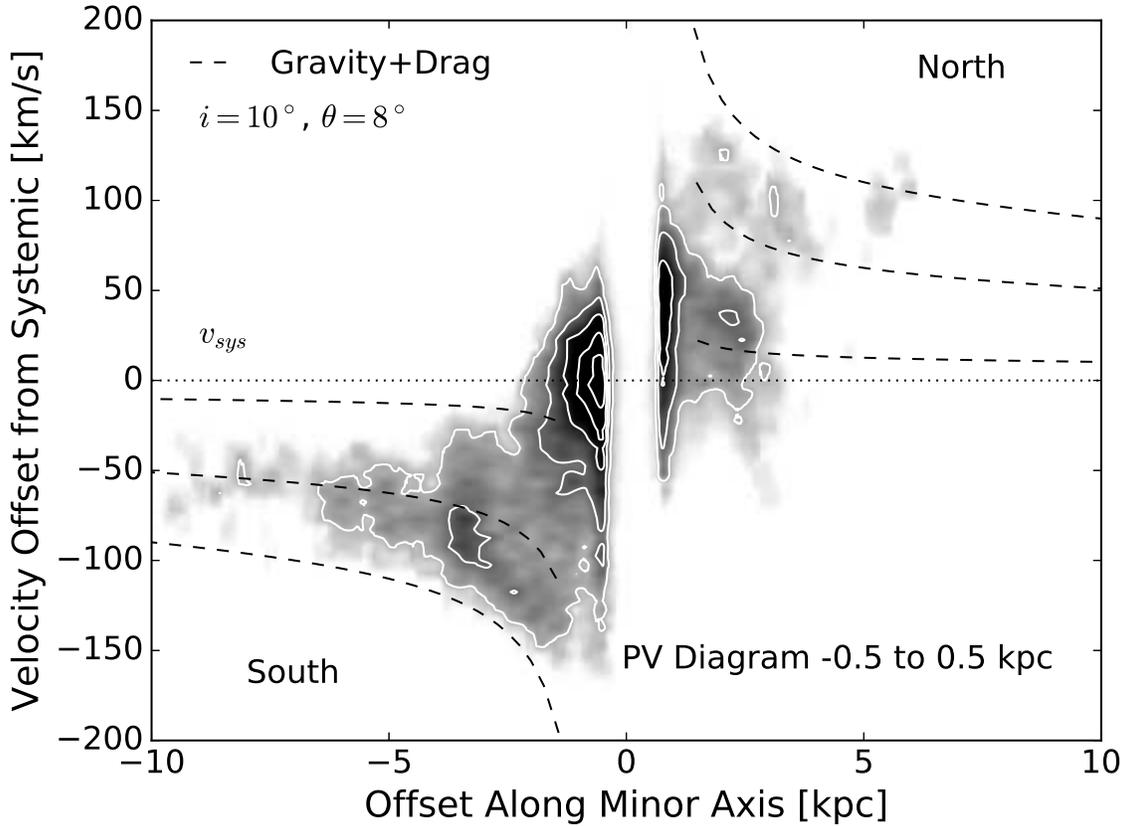} 
\caption{Same as Figure~\ref{fig:pv1} except with a model that includes both gravitational and drag forces on the \hi\ gas. \label{fig:pvgd}}
\end{figure*}

\subsection{Minor Axis Velocity Decrease}

\begin{figure}[ht]
\plotone{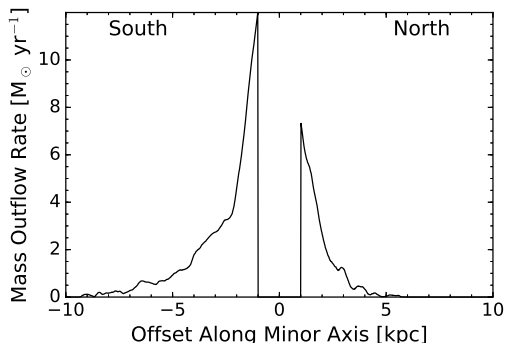} 
\caption{Integrated mass outflow rate within $\pm 0.5$ kpc of the minor axis derived from the surface density shown in Figure~\ref{fig:nhi} and the velocity profile of the best-fit model. The central $\pm 1$ kpc is masked as there is significant \hi\ absorption near the midplane. \label{fig:mdot}}
\end{figure}

The most remarkable feature of the minor axis \hi\ kinematics is the decrease of the typical velocity amplitude from about $100\,\kms$ to $50\,\kms$ over the projected distance range from about 1.5 kpc to 10 kpc (Figure~\ref{fig:pv1}). This decline is very obvious along the minor axis toward the South. The kinematics to the North are broadly consistent with this same velocity profile, although the \hi\ intensity is substantially less and the gas is only detected to about 5 kpc in projection. 

The velocity amplitude decrease from 1.5 to 10 kpc is not consistent with \hi\ clouds that have been accelerated by the wind and then launched on ballistic trajectories, and consequently demonstrate that an additional force is required. This is because the inferred outflow velocities of these clouds is on order $500-600\,\kms$ (for $80^\circ$ disk inclination) at a projected height of 1.5 kpc and the mass of M82 is insufficient to slow the clouds by 50\% from 1.5 kpc to 10 kpc. This point is illustrated by the dashed lines in Figure~\ref{fig:pv1}. The lines show the projected velocity of a test particle calculated with \texttt{galpy} and the mass model for M82 developed in \S\ref{sec:mass}. The middle line shows the trajectory for a test particle on the symmetry axis of the bicone (at an angle of $10^\circ$ relative to the plane of the sky) and the other lines show test particles at $\pm8^\circ$ on the nominal edges of a cone with a $16^\circ$ opening angle. The \hi\ kinematics clearly decrease much faster than expected from gravitational forces alone. They are also inconsistent with cooling of the \hi\ gas from the hot wind. \hi\ produced from cooling in a hot wind should have the velocity of the hot wind, which \citet{strickland09} estimate to be as high as $\sim 2000\,\kms$. The \hi\ kinematics also demonstrate that the cold \hi\ clouds do not experience continual acceleration, nor even free expansion into the halo of M82. 

One possible explanation for this decrease in projected velocity is that the inclination of the bicone is substantially different from the stellar disk, as this would decrease the \hi\ outflow velocity. We find that the \hi\ velocities do diminish by the required amount from $1.5 - 5$ kpc if the symmetry axis of the cone is tilted toward us by $30^\circ$ relative to the plane of the sky. However, in this case the \hi\ velocities continue to decrease to the systemic velocity, and do not reproduce the relatively flat velocity profile seen from $5-10$ kpc. Other possible explanations include that the \hi\ experiences some additional drag force, the \hi\ does not represent a continuous stream of material, but instead is forming in situ (although not from the hot wind), and that the \hi\ kinematics have been substantially affected by tidal interactions. 

If the \hi\ consists of clouds flowing outward along the edge of the bicone, they could experience a drag force if they encounter ambient material outside of the bicone. This material could be halo gas of M82 that is not part of the wind, or tidal material that is approximately stationary relative to the wind. We model the drag force with the drag equation $f_{drag} = C_d \rho_{amb} v^2 A$, where $C_d$ is the drag coefficient, $\rho_{amb}$ is the density of the ambient medium, $v$ is the velocity of the \hi\ relative to this medium, and $A$ is the cross section of a typical \hi\ cloud. We then make the assumption that $C_d \rho_{amb} A \propto r^{-2}$. This would be true if $C_d A$ were constant and $\rho_{amb} \propto r^{-2}$. The coefficient $C_d$ is approximately constant for spheres (of constant cross section $A$) in a supersonic flow, while $\rho_{amb} \propto r^{-2}$ is characteristic of mass outflow at a constant rate at constant velocity. Based on measurements by \citet{leroy15}, we expect the density profile is somewhat steeper with $\rho_{amb} \propto r^{-3 \rightarrow -4}$ for the dust and molecular gas\footnote{While \citet{leroy15} did measure  $r^{-2}$ for \hi, that is more due to the abundance of tidal material, than because the wind material follows $r^{-2}$.}. To maintain our assumption in this case would require the drag coefficient to increase with $r$, which could happen if the Reynolds number decreased with radius. We find that the addition of a drag force with $C_d \rho_{amb} A \propto r^{-2}$ provides a reasonable match to the data. This model is shown in Figure~\ref{fig:pvgd}. 

If the velocity change of the \hi\ gas is due to a drag force imposed by the ambient medium, the non-gravitational deceleration of the test particles that best match the data constrain the product of $C_d \rho_{amb} A_{cl}$. If we assume these are spherical clouds of \hi\ gas with uniform density $\rho_{cl}$ and radius $R_{cl}$, the non-gravitational deceleration is: 
\begin{equation}
    \frac{dv}{dt} = \frac{3}{4} C_d \left( \frac{\rho_{amb}}{\rho_{cl}} \right) \frac{v_{cl}^2}{R_{cl}}
\end{equation}
At $2$ kpc from the midplane, the deprojected \hi\ velocity is about $500\,\kms$ and the best-fit model has a non-gravitational deceleration of $1.8\times10^{-7} {\rm cm\,s}^{-2}$. For $R_{cl} = 10$ pc, the density ratio is $\sim 0.0044 (0.5/C_d)$. If the cloud particle density is $10\,{\rm cm}^{-3}$, the ambient medium at $2$ kpc from the midplane must be relatively dense with $0.044\,{\rm cm}^{-3}$. This is substantially denser than the hot wind, although the wind has much larger velocities so it is unlikely to be the same medium. If drag forces decelerate the \hi\ gas, then the ambient medium responsible for the deceleration must be distinct from the hot wind, and at higher density than the typical hot, coronal gas of galactic halos. This could be due to material stripped out of the disk by the tidal interaction with M81. The tidal interaction may also be the main origin of the decrease in the \hi\ velocity, rather than a drag force due to the ambient medium. 

The combination $\rho_{cl} R_{cl}$ in Equation~1 is the column density of the clouds in the outflow. If the majority of the material in the cloud is \hi, then this quantity can be measured with \hi\ emission maps or UV/visible absorption line studies with sufficiently high resolution and sensitivity. Such a measurement, combined with the acceleration and the velocity of the clouds, could then be used to solve for the ambient density required to produce the observed deceleration. The angular size of a $\sim 10-20$ pc cloud at the distance of M82 is $0.5-1\arcsec$. This is a challenge for current \hi\ facilities, but may be feasible with UV/visible absorption line studies toward background sources.

Similar to \citet{leroy15}, we have used the mass distribution and velocity information to derive the mass flux associated with the \hi\ emission. Under the assumption that all of the \hi\ is associated with an outflow, Figure~\ref{fig:mdot} shows the mass flux computed from the product of the mass surface density (Figure~\ref{fig:nhi}), the velocity from our outflow model, and the $\pm 0.5$ kpc width of the central region of our position-velocity diagram. The main differences from the calculation and figure shown in \citet{leroy15} are that we used a smaller region width (1 kpc, rather than 3 kpc) and we used the velocity profile, rather than a fixed outflow speed of $450\,\kms$. While the mass flux we have derived near the midplane is lower than in \citet{leroy15}, largely because of the smaller region width, the mass flux is detected out to about twice the distance from the midplane, although it drops to a fraction of a solar mass per year beyond 4 kpc. This is also approximately the point where the mass flux is comparable to the outflow rate computed by \citet{strickland09} for the hot gas. This \hi\ outflow rate is about an order of magnitude less than the star formation rate. 

The \hi\ velocity decrease along the minor axis may be so substantial that the \hi\ will not escape the halo. At a projected height of 10 kpc, the line of sight velocity has dropped to $50\,\kms$ relative to the systemic velocity, which corresponds to an inclination-corrected outflow velocity of about $300\,\kms$. This is sufficiently comparable to the escape velocity of our mass model that the ultimate fate of the gas will depend on assumptions in our model for the drag force on the cold gas, the impact of tidal forces in the M81 group, and the halo model for M82. Within the context of our model, the \hi\ would finally stall at 70--80 kpc after about 500 Myr. The \hi\ gas could consequently form a cold ``fountain'' as proposed by \citet{leroy15}, rather than substantially add to the intergalactic medium. We searched in our data for any evidence of material that is falling back, but do not see any. This included a careful inspection of the non-masked data, which has somewhat greater sensitivity, although at the expense of more artifacts. 

\subsection{Biconical Frustum} \label{sec:frustum} 

\begin{figure*}[ht]
\plotone{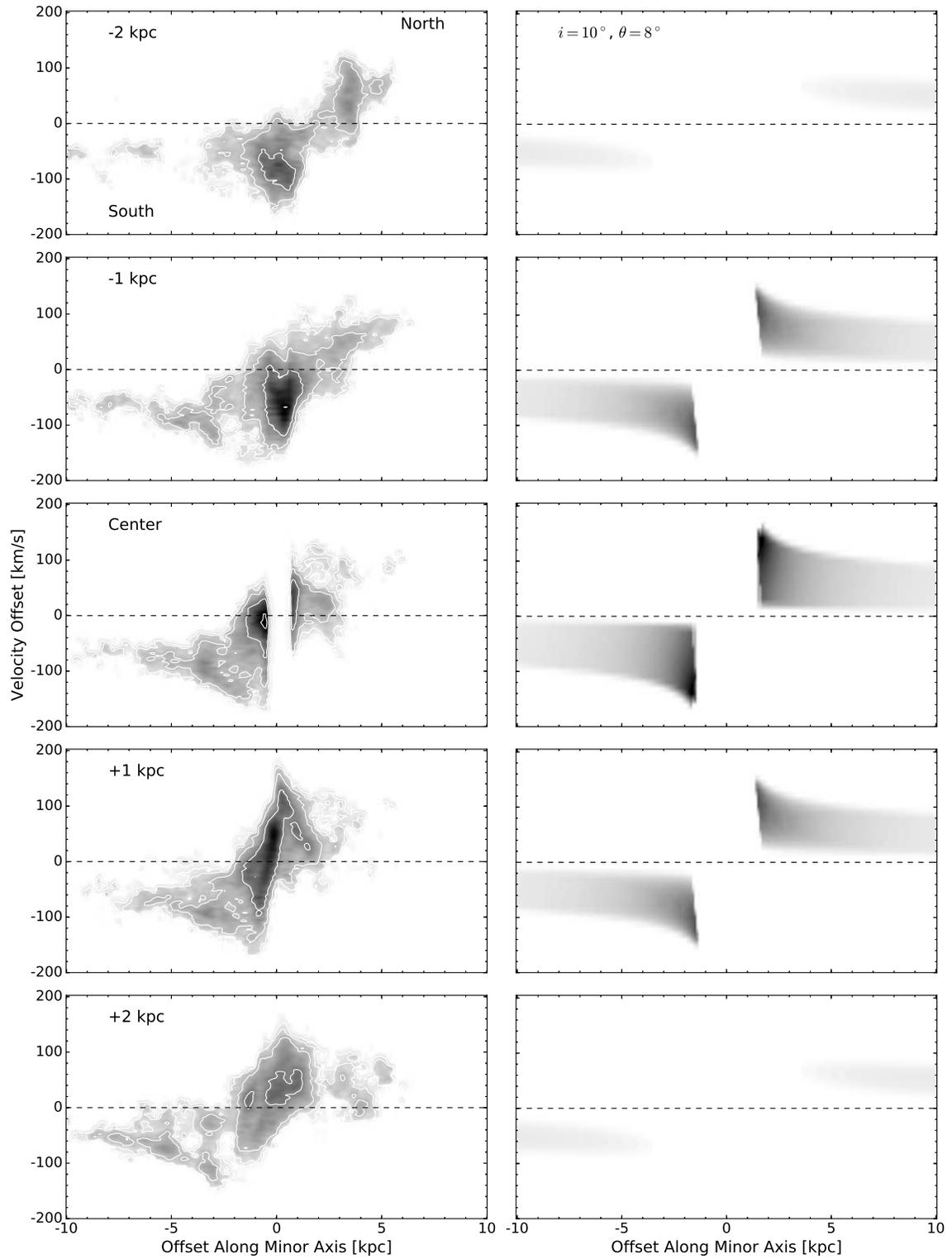} 
\caption{Five position-velocity diagrams offset for the observed \hi\ gas ({\it left}) and a model biconical frustum outflow with an R=1kpc base on the major axis ({\it right}). Each row shows a 1\,kpc wide slice perpendicular to the major axis. The \hi\ data show the five positions marked in Figure~\ref{fig:intvel}, while the model shows the projected position-velocity diagram of a model filled biconical frustum with $i = 10^\circ$ and opening angle $\theta = 8^\circ$ (half width). The model includes both the gravitational deceleration and the drag force shown in Figure~\ref{fig:pvgd}. \label{fig:pvall}}
\end{figure*}

A simple bicone model is not a good match to the spatial and velocity distribution of the \hi\ gas. Figures~\ref{fig:intvel} and \ref{fig:channels} clearly show the \hi\ gas does not have a conical shape. The opening angle derived from line splitting in the cold gas by \citet{leroy15} is also smaller than values derived closer to the plane from the warm, ionized gas \citep{heckman90,mckeith95}, which suggests the cold gas is more columnated. Another argument against a simple bicone is the presence of outflowing material at offset positions parallel to the minor axis. Figure~\ref{fig:intvel} shows five slices parallel to the minor axis that are 1 kpc wide and offset at $+2, +1, 0, -1$, and $-2$ kpc along the midplane. Figure~\ref{fig:pvall} shows the position velocity at each of those five positions. This figure clearly shows the \hi\ velocity profile observed within $\pm 0.5$ kpc of the minor axis (Center) is also seen in positions offset by as much as $\pm2$ kpc, albeit at somewhat lower intensity. A simple bicone with an opening angle of $8^\circ$ (half width) would have no flux within about $\pm 7$ kpc of the major axis in the offset positions at $\pm 2$ kpc. While the intensity at these offset positions could be explained with a larger opening angle, a larger opening angle would also produce a much larger range of velocities at each position that would not be consistent with the observations. 

We have consequently compared the data to a biconical frustum, rather than a simple bicone. Since a frustum is a conical shape with the top removed, we use this representation to model the origin of the outflow from a broader region of the disk than the single point of a cone.  This shape is a better physical match to the spatial extent of the starburst, yet maintains the relatively narrow range of velocities observed further from the midplane \citep[see][for the application of a frustum to NGC 253]{westmoquette11}. We find that a frustum radius of 1 kpc at the midplane and an opening angle half width of $8^\circ$ is a reasonable match to the data. The outline of this biconical frustum is shown in Figure~\ref{fig:intvel}, and position-velocity diagrams for this model are shown on the right side of Figure~\ref{fig:pvall} at the same five offset positions as the data on the left. The choice of a 1 kpc radius at the midplane was motivated by the physical extent of the starburst region \citep[$\sim 500$ pc][]{forster03}, but increased to provide a better match to the observed position-velocity diagrams. The velocity of the gas in the model was calculated at each point based on the radial distance from the center of the model. The velocity includes gravity and drag forces as in Figure~\ref{fig:pvgd}, and was normalized such that the initial velocity of the gas matches the observed \hi\ kinematics at 1.5 kpc from the midplane. The model does not include any material closer than 1.5 kpc from the center, as we assumed this region is substantially affected by the disk kinematics. The main disagreement between the model and data are that the model has a somewhat broader velocity distribution. This could be because our simple model adopts a uniformly filled frustum. The model would better match the data if the material were more concentrated on the symmetry axis, although physically that would be in conflict with the expected location of the hot wind. At larger distances, the density is assumed to decrease as $n \propto r^{-2}$. 

\subsection{Alternative Scenarios}

It is also possible that the \hi\ gas does not represent a continuous population of clouds flowing from the disk, but is instead partially or completely produced in situ. In this case, one potential origin is the dissociation of molecular gas. The molecular gas measurements from \citet{leroy15} within a few kpc of the midplane show similar kinematics as the atomic gas; however, the mass density of molecular gas is substantially lower than the \hi\ off the midplane, and decreases much more rapidly with distance from M82. Deeper molecular gas observations would help to further test this hypothesis, along with calculations of the disassociation rate. 

Another possibility is that the \hi\ gas has radiatively cooled from the hot phase \citep{wang95a,wang95b}. The basic physical picture, which was developed in detail by \citet{thompson16}, starts with cool gas clouds in the disk that are initially accelerated and shredded by the hot wind. The addition of this cloud material then increases the mass loading in the wind, and then this additional mass seeds thermal and radiative instabilities that precipitate cool gas out in the wind at larger distances. Following the wind model of \citet{chevalier85}, a key dimensionless parameter is the mass loading parameter $\beta$, which describes the ratio of the mass outflow rate $\dot{M}_W$ in the wind to the star formation rate $\dot{M}_*$, that is $\dot{M}_{W} = \beta \dot{M}_*$. The radiative cooling requires some minimum mass loading due to the shape of the cooling function. For the case of M82,  \citet{hoopes03} find weak evidence of cooling on large scales and \citet{strickland09} estimate a fairly low mass loading factor of $\beta = 0.2 - 0.5$ in the core, although the wind could be strongly mass loaded on larger scales. \citet{thompson16} therefore conclude that their model may not apply to M82, and note the additional complication of tidal material. The monotonic decline in the characteristic \hi\ velocity with projected distance also does not seem consistent with radiative cooling  because the velocity profile is inconsistent with the inferred velocity profile for the hot wind.

There is some phenomenological similarity between the velocity decrease in M82 on kpc scales and the ionized gas kinematics observed in nearby AGN on scales approximately an order of magnitude smaller. Studies of Seyfert galaxies such as NGC 1068 \citep{crenshaw00}, Mrk 3 \citep{ruiz01}, and NGC 4151 \citep{das05} are reasonably well matched with constant acceleration to 100-300 pc, followed by constant deceleration to the systemic velocity. \citet{everett07} explored Parker wind models to explain the $\sim 100$ pc outflows in AGN, but found no self-consistent temperature profile that could explain all of the kinematics. They did note that the generic hydromagnetic winds explored by \citet{matzner99} could explain the acceleration, and there are morphological similarities between that model and the observations of \citet{may17}, but it was not clear how to explain the deceleration without interaction with the surrounding medium.  

\section{Summary} \label{sec:sum}

We have presented new VLA observations of the archetypical starburst galaxy M82, and combined them with earlier observations from the GBT. These data trace \hi\ emission out to a projected distance of 10 kpc from the center of the galaxy with an angular resolution of $\approx 24\arcsec$. The velocity amplitude to both the North and South decreases from about $140 \rightarrow 50\,\kms$ over a projected distance of $1.5 \rightarrow 5$ kpc from the midplane. The \hi\ intensity is substantially greater to the South, where \hi\ is detected out to a projected distance of 10 kpc. 

These minor axis \hi\ kinematics are not consistent with continual acceleration of the atomic gas from the midplane, nor with gas clouds launched on approximately ballistic trajectories. The \hi\ gas is also three to four times slower than the inferred velocity of the hot superwind fluid traced by X-ray data, and therefore the \hi\ gas is not consistent with material that has cooled from that phase. The \hi\ kinematics are more similar to the molecular and warm, ionized gas that appears to trace the outer sheath of the superwind, although the warm, ionized gas has only been measured closer to the midplane. 

We use a new mass model for M82 to demonstrate that an additional force is required to slow the \hi\ gas, and that the observed velocity profile is consistent with drag caused by the ambient environment. This requires that the \hi\ gas is spatially distinct from the hot superwind fluid, and is not just the cool tail of the temperature distribution. The magnitude of the required drag force constrains the density ratio of the clouds and ambient medium, as well as the typical cloud size, which could be tested with future, higher resolution \hi\ observations. The high velocities and density contrast between the outflow and ambient medium would also produce shock-heated gas, which may contribute to the X-ray emission. 

One alternate possibility is that the \hi\ gas is not a continuous stream of material flowing from the disk, but rather is partially or completely produced in situ by dissociation of the molecular gas and/or adiabatic cooling of the warm, ionized gas that forms the interface between the superwind and the ambient medium. We disfavor this scenario because it does not naturally explain the systematic decrease in the \hi\ velocities with distance from the midplane, at least not without a similar drag force on that material. There is also a very substantial velocity contrast between the superwind and the other phases. 
Deeper observations of the molecular and especially the warm ionized gas that extend further from the midplane would help reveal the extent to which that material shares the same kinematic profile as the \hi\ gas.

A final possibility is that the \hi\ along the minor axis is largely tidal debris viewed in projection near the minor axis. Such a superposition would be an unfortunate coincidence, and does not appear consistent with either the symmetric kinematic profile about the midplane or the larger-scale tidal streams of \hi. Deeper, wider-field \hi\ observations, combined with new models of the interactions in the M81 group, could help to quantify the contribution of \hi\ tidal streams, if any, to the \hi\ observed along the minor axis.

The ultimate fate of the \hi\ gas is unclear. Our detection of the velocity decrease along the minor axis, combined with our new mass model for M82, demonstrates that there is likely some additional drag force that slows the cold phase of the wind. This velocity decrease lowers the wind speed to be comparable to the escape velocity from the halo, and thus the ultimate fate of the gas could be fallback onto the disk in a cold fountain on $\sim100$ Myr timescales, long-term residence in the circumgalactic medium, or escape into the intergalactic medium. How the gas is distributed among these three scenarios depends on any additional drag forces on the wind at larger distances, details of the halo mass distribution, and the tidal field of the M81 group. 

\acknowledgments

The National Radio Astronomy Observatory is a facility of the National Science Foundation operated under cooperative agreement by Associated Universities, Inc. We thank Todd Thompson for discussions of Galactic winds and outflows. We also thank Norm Murray for his helpful referee report. The work of PM is partially supported by the National Science Foundation under Grant 1615553 and by the Department of Energy under Grant DE-SC0015525. The work of AKL is partially supported by the National Science Foundation under Grants No. 1615105, 1615109, and 1653300. ADB acknowledges support from the National Science Foundation grant AST-1412419. 

Facilities: \facility{VLA}, \facility{GBT}.

\end{document}